\documentclass[aps,prd,twocolumn,showpacs,superscriptaddress,preprintnumbers]{revtex4-1}

\usepackage{graphicx}
\usepackage{latexsym,amsmath,amsfonts,amssymb}
\usepackage{hyperref}
\usepackage{bbm}
\pdfoutput=1

\def\BR{\mathbb{R}}

\makeatother

\begin{document}

\preprint{IPMU-14-0322, UT-14-43}
\title{Holographic Interpolation between $a$ and $F$}

\author{Teruhiko Kawano}
\affiliation{
Department of Physics, Faculty of Science,
The University of Tokyo, \\
Bunkyo-ku, Tokyo 113-0033, Japan
}
\author{Yuki Nakaguchi} 
\affiliation{
Department of Physics, Faculty of Science,
The University of Tokyo, \\
Bunkyo-ku, Tokyo 113-0033, Japan
}
\affiliation{
Kavli Institute for the Physics and Mathematics of the Universe (WPI), The University of Tokyo,\\
5-1-5 Kashiwa-no-Ha, Kashiwa City, Chiba 277-8568, Japan
}

\author{Tatsuma Nishioka}
\affiliation{
Department of Physics, Faculty of Science,
The University of Tokyo, \\
Bunkyo-ku, Tokyo 113-0033, Japan
}

\begin{abstract}
An interpolating function $\tilde F$ between the $a$-anomaly coefficient in even dimensions and the free energy on an odd-dimensional sphere has been proposed recently and is conjectured to monotonically decrease along any renormalization group flow in continuous dimension $d$.
We examine $\tilde F$ in the large-$N$ CFT's in $d$ dimensions holographically described by the Einstein-Hilbert gravity in the AdS$_{d+1}$ space.
We show that $\tilde F$ is a smooth function of $d$ and correctly interpolates the $a$ coefficients and the free energies.
The monotonicity of $\tilde F$ along an RG flow follows from the analytic continuation of the holographic $c$-theorem to continuous $d$, which completes the proof of the conjecture.
\end{abstract}

\date{\today}

\maketitle

\section{Introduction}
A measure of degrees of freedom in a quantum field theory (QFT) remains to be elucidated in arbitrary $d$ dimensions.
Physically, it decreases monotonically as the energy scale is lowered because of the decoupling of massive particles.
Implementation of such a measure in any QFT in diverse dimensions is intriguing and desirable to characterize the behavior under a renormalization group (RG) flow.

For even $d$, the conformal anomaly in the stress-energy tensor \footnote{We define the stress-energy tensor by $T_{\mu\nu} \equiv \frac{2}{\sqrt{g}} \frac{\delta I}{\delta g^{\mu\nu}}$ for an action $I$. The Euler density is normalized to be $\int_{S^d} d^d x \sqrt{g}\, E_d = 2$.} 
\begin{align}\label{WeylAnomaly}
\langle T_\mu^{~\mu} \rangle = \frac{(-1)^{\frac{d}{2}+1}}{2} a\, E_d + \sum_{i} b_i\, I_i \ ,
\end{align}
defines the unique $a$ coefficient for the Euler density $E_d$ and several $b_i$ coefficients for the Weyl invariants $I_i$ labeled by an integer $i$.
The $a$ coefficients are believed to be monotonically decreasing along any RG flow, namely the value $a_\text{UV}$ at the ultra-violet (UV) fixed point is equal or greater than that $a_\text{IR}$ at the infra-red (IR) fixed point,  $a_\text{UV}\ge a_\text{IR}$.
This statement was established in two dimensions by the Zamolodchikov's $c$-theorem \cite{Zamolodchikov:1986gt} and in four dimensions by the $a$-theorem  \cite{Cardy:1988cwa,Komargodski:2011vj,Komargodski:2011xv}.
On the other hand, the $F$-theorem asserts that the free energy, $F\equiv (-1)^{\frac{d-1}{2}}\log Z_{S^d}$, defined by the conformal invariant partition function $Z_{S^d}$ on $S^d$ of radius $R$, decreases under any RG flow in odd dimensions \cite{Jafferis:2011zi,Klebanov:2011gs}.
A proof for $d=3$ was presented by \cite{Casini:2012ei} through the relation of the free energy to the entanglement entropy $S$ across an entangling surface $S^{d-2}$ of radius $R$ in $\BR^{1,d-1}$ \cite{Casini:2011kv}
\begin{align}\label{EEandF}
F = (-1)^{\frac{d-1}{2}} S \ ,
\end{align}
that holds for odd $d$ up to UV divergences.

These two proposals look quite different at first sight, but share the fact that both the $a$ coefficient and the free energy can be read off on $S^d$;
the former arises from the integration of the trace of the stress-energy tensor \eqref{WeylAnomaly} and the latter from the partition function.
To interpolate between the $a$ coefficient and the free energy, Giombi and Klebanov define a new function \cite{Giombi:2014xxa}
\begin{align}\label{Ftilde}
\tilde F \equiv \sin\left( \frac{\pi d}{2}\right) \log Z_{S^d} \ ,
\end{align}
which correctly reduces to the free energies for odd $d$.
They show as $d$ approaches to even integers \footnote{
There is no sign factor $(-1)^{d/2}$ in the right hand side because
our convention of the $a$-anomaly \eqref{WeylAnomaly} differs from theirs in \cite{Giombi:2014xxa}.} (see also \cite{Yonekura:2012kb} as a related work)
\begin{align}\label{FandA}
\tilde F = \frac{\pi}{2}\,a \ .
\end{align}
Note that the partition function $Z_{S^d}$ used in \eqref{Ftilde} is conformal invariant and UV divergent for even $d$.
The relation \eqref{FandA} follows from the fact that the conformal invariant partition function in $d = 2n + \epsilon$ dimensions behaves as $\log Z_{S^d} = (-1)^{\frac{d}{2}}\frac{a}{2\epsilon}+O(1)$ for small $\epsilon$. 
This is because one has to add a local counter term 
\begin{align}
I_\text{c.t.} = (-1)^{\frac{d}{2}+1}\frac{a}{2\epsilon} \int_{S^d} d^d x\sqrt{g}\, E_{2n} \ ,
\end{align}
to the partition function to obtain the renormalized partition function $\log Z^\text{(ren)}_{S^d} = \log Z_{S^d} + I_\text{c.t}$, reproducing the conformal anomaly $\log Z^\text{(ren)}_{S^{2n}} = (-1)^{n+1} a \log R$ on $S^{2n}$ of radius $R$ in $\epsilon\to 0$ limit.

The function $\tilde F$ is also defined for non-integer $d$ and therefore smoothly interpolates between the $a$ coefficients in even dimensions and the free energies in odd dimensions.
They conjecture that $\tilde F$ is positive and decreases along any RG flow in arbitrary $d$ dimensions, based on several examples including a double-trace deformation of the large-$N$ conformal field theory (CFT).
We will call their proposal the $\tilde F$-theorem.

In this letter, we provide a further evidence to the $\tilde F$-theorem from the holographic viewpoint.
To this end, we take advantage of the relation \eqref{EEandF} and calculate the holographic entanglement entropy \cite{Ryu:2006bv,Ryu:2006ef} across a sphere $S^{d-2}$ in the Einstein-Hilbert gravity on the AdS$_{d+1}$ space.
We perform the dimensional regularization in the bulk and obtain the analytic result of $\tilde F$ that is a positive and smooth function of dimension $d$.
We show that the equality \eqref{FandA} holds for even $d$ and furthermore prove the $\tilde F$-theorem that follows from the holographic $c$-theorem \cite{Girardello:1998pd,Freedman:1999gp,Myers:2010xs,Myers:2010tj} assuming the dimensional continuation of the null energy condition.

\section{Holographic proof of the $\tilde F$-theorem}
We will evaluate $\tilde F$ with the relation \eqref{EEandF} between the free energy on $S^d$ and the entanglement entropy across $S^{d-2}$.
The latter can be holographically calculated by the Ryu-Takayanagi formula in the Einstein-Hilbert gravity \cite{Ryu:2006bv,Ryu:2006ef}
\begin{align}
S = \frac{\text{Area}(\gamma)}{4G_N^{(d+1)} }\ ,
\end{align}
where $G_N^{(d+1)}$ is the Newton constant, and $\gamma$ stands for the $(d-1)$-dimensional minimal surface in the AdS$_{d+1}$ space, whose boundary is the entangling surface $S^{d-2}$.
Since the boundary of the AdS$_{d+1}$ space is the flat space $\BR^{1,d-1}$, we will use the Poincar{\'e} coordinates 
\begin{align}
ds^2 = L^2 \frac{dz^2 - dt^2 + dr^2 + r^2 d\Omega_{d-2}^2}{z^2} \ ,
\end{align}
where $L$ is the AdS radius.
The entangling surface is located at $t=0$ and $r=R$ at the boundary $z=0$.
In these coordinates, the minimal surface $\gamma$ in the bulk is a hemi-hypersphere satisfying $r^2 + z^2 = R^2$ \cite{Ryu:2006bv,Ryu:2006ef}.
This solution leads the entanglement entropy across $S^{d-2}$
\begin{align}
S = \frac{1}{4G_N^{(d+1)}} L^{d-1}\text{Vol}(S^{d-2}) \int_{\epsilon/R}^1 dy \frac{(1-y^2)^{\frac{d-3}{2}}}{y^{d-1}} \ , 
\end{align}
where we introduced a small cutoff at $z=\epsilon$ to regularize the UV divergence and $\text{Vol}(S^{d-2})$ is the volume of a unit $(d-2)$-dimensional round sphere.
Expanding the integrand with respect to $y$ and performing the integration, one obtains the UV divergent parts of the entanglement entropy.
We, however, want to employ the dimensional regularization instead of putting the UV cutoff at $z=\epsilon$ for our purpose.
So we take $\epsilon=0$ and carry out the integral in the range $1 < d < 2$, that yields
\begin{align}\label{HEE}
S = \frac{L^{d-1}}{4G_N^{(d+1)}} \pi^{\frac{d}{2} -1} \Gamma \left( 1-\frac{d}{2}\right) \ .
\end{align}
Then we analytically continue $d$ to any real value.
It is clear that there are poles at even $d$ in the entanglement entropy \eqref{HEE} corresponding to the conformal anomalies.
Finally, using the relations \eqref{EEandF} and \eqref{Ftilde}, and the formula $\Gamma(z) \Gamma(1-z) = \pi/\sin(\pi z)$, we obtain $\tilde F$ in the holographic theories
\begin{align}\label{HolographicFtilde}
\tilde F = \frac{L^{d-1}}{4G_N^{(d+1)}} \frac{\pi^\frac{d}{2}}{\Gamma\left( \frac{d}{2}\right)} \ .
\end{align}
This is manifestly a positive and smooth function of dimension $d$ without poles at even $d$.

Now let us extrapolate the holographic values of $\tilde F$ to even dimensions and see if the relation \eqref{FandA} holds.
The $a$ coefficients holographically computed in the Einstein-Hilbert gravity are known to be
\cite{Imbimbo:1999bj,Schwimmer:2008yh,Myers:2010xs,Myers:2010tj}
\begin{align}\label{a_Holographic}
a = \frac{L^{d-1}}{2\pi G_N^{(d+1)}}\frac{\pi^\frac{d}{2}}{\Gamma\left( \frac{d}{2}\right)} \ .
\end{align}
Combining it with \eqref{HolographicFtilde}, we confirm the relation \eqref{FandA} between $\tilde F$ and $a$.
Moreover, imposing the null energy condition in the bulk, the holographic $c$-theorem states that the $a$ coefficient given by \eqref{a_Holographic} satisfies the monotonicity, $a_\text{UV} \ge a_\text{IR}$, for positive integer $d$ \cite{Girardello:1998pd,Freedman:1999gp,Myers:2010xs,Myers:2010tj}.
Assuming the analytic continuation of dimension $d$ in the gravity, the holographic $c$-theorem holds for $d\ge 1$ \footnote{The null energy condition $T_{\mu\nu}\xi^\mu\xi^\nu \ge 0$ is crucial in the proof of the holographic $c$-theorem \cite{Freedman:1999gp,Myers:2010xs,Myers:2010tj} where the $d$-dimensional null vector $\xi$ has only two non-zero components $\xi^z$ and $\xi^t$.
Thus defining a formal null vector $\xi=(\xi^z, \xi^t, 0, \cdots, 0)$ in continuous $d$ dimensions, the proof can be carried over for $d\ge 1$.
} which assures the $\tilde F$-theorem due to the relation \eqref{FandA}.

\begin{acknowledgments}
We are grateful to Y.\,Tachikawa and K.\,Yonekura for valuable discussions and to S.\,Giombi and I.\,Klebanov for correspondence. The work of T.\,K. was supported in part by 
a Grant-in-Aid \#23540286 from the MEXT of Japan.
The work of Y.\,N. was supported in part by JSPS Research Fellowship for Young Scientists and World Premier International Research Center Initiative (WPI) from the MEXT of Japan.
\end{acknowledgments}



\bibliographystyle{apsrev4-1}
\bibliography{Holographic_Ftilde}

\end{document}